\documentclass[conference,a4paper]{IEEEtran}
\IEEEoverridecommandlockouts

\usepackage[draft]{graphicx}
\usepackage{balance}
\usepackage[T1]{fontenc}
\usepackage{ifthen}
\usepackage[cmex10]{amsmath} 

\usepackage[lined,boxed,commentsnumbered,linesnumbered, ruled]{algorithm2e}
\usepackage[url,hyperrefblack,notheorems,IEEEtran]{research17}
\usepackage{comment}
\usepackage{booktabs} 
\usepackage{mathtools}
\usepackage{amsmath,amssymb,amsfonts,amsthm}
\usepackage{tikz} 
\tikzset{font={\fontsize{10pt}{12}\selectfont}}
\usetikzlibrary{shapes, patterns,decorations.text, decorations.pathreplacing,plotmarks}

\newtheorem{lemma}{\mylemmaname}
\newtheorem{theorem}{\mytheoremname}
\newtheorem{definition}{\mydefinitionname}

\newtheorem{remark}{\myremarkname}
\newtheorem{claim}{\myclaimname}

\usepackage[capitalize]{cleveref}

\crefname{equation}{\unskip}{\unskip}
\crefname{claim}{Claim}{Claims} 
\usepackage{array}
\usepackage{multirow}
\newcolumntype{C}[1]{>{\centering\arraybackslash}p{#1}}
\allowdisplaybreaks

\makeatletter
\def\@xfootnote[#1]{%
	\protected@xdef\@thefnmark{#1}%
	\@footnotemark\@footnotetext}
\makeatother

\renewcommand{\vect}[1]{\vectg{#1}} 
\renewcommand{\vmat}[1]{\bm{\mat{#1}}} 
\newcommand{\code}[1]{\mathscr{#1}} 
\newcommand{\collect}[1]{\mathscr{#1}} 
\newcommand*{\Resize}[2][4]{\resizebox{#1}{!}{\ensuremath{#2}}} 
\makeatletter
\renewcommand*\env@matrix[1][*\c@MaxMatrixCols c]{%
  \hskip -\arraycolsep
  \let\@ifnextchar\new@ifnextchar
  \array{#1}}
\makeatother

\newcommand{\HP}[1]{\HH\left(#1\right)} 
 
\newcommand{\bigHP}[1]{\HH\bigl(#1\bigr)}

\newcommand{\HPcond}[2]{\HH\left(#1 \kern0.1em\middle|\kern0.1em #2\right)}
\newcommand{\eHPcond}[2]{\HH(#1 \kern0.1em|\kern0.1em #2)} 
\newcommand{\bigHPcond}[2]{\HH\bigl(#1 \kern-0.1em \bigm| \kern-0.1em#2\bigr)}
\newcommand{\BigHPcond}[2]{\HH\Bigl(#1 \kern-0.1em \Bigm| \kern-0.1em#2\Bigr)}
\newcommand{\MI}[2]{\II\left(#1 \kern0.1em{;}\kern0.1em #2\right)} 
\newcommand{\eMI}[2]{\II(#1 \kern0.1em{;}\kern0.1em #2)} 
\newcommand{\bigMI}[2]{\II\bigl(#1 \kern0.1em{;}\kern0.1em #2\bigr)}
\newcommand{\BigMI}[2]{\II\Bigl(#1 \kern0.1em{;}\kern0.1em #2\Bigr)}
\newcommand{\MIcond}[3]{\II\left(#1 \kern0.1em{;}\kern0.1em #2 \kern0.1em\middle|\kern0.1em #3\right)}
\newcommand{\eMIcond}[3]{\II(#1 \kern0.1em{;}\kern0.1em #2 \kern0.1em|\kern0.1em #3)} 
\newcommand{\bigMIcond}[3]{\II\bigl(#1 \kern0.1em{;}\kern0.1em #2 \kern-0.1em \bigm| \kern-0.1em#3\bigr)}
\newcommand{\BigMIcond}[3]{\II\Bigl(#1 \kern0.1em{;}\kern0.1em #2 \kern-0.1em \Bigm| \kern-0.1em#3\Bigr)}
\newcommand{\m}{\color{magenta}} %
\usepackage{soul} 

\DeclareSymbolFont{matha}{OML}{txmi}{m}{it}
\DeclareMathSymbol{\varv}{\mathord}{matha}{118}
\usepackage{upgreek}
\usepackage{tablefootnote}
\usepackage{balance}

\begin{document}
%
\sloppy \title{Single-Server Pliable Private Information \\Retrieval With Side Information}
%

\author{
    \IEEEauthorblockN{Sarah A.~Obead, Hsuan-Yin Lin, and Eirik Rosnes}
    \IEEEauthorblockA{Simula UiB, N--5006 Bergen, Norway}
    \IEEEauthorblockA{Emails: \{sarah, lin, eirikrosnes\}{@}simula.no  }
}

\maketitle

\begin{abstract}
   We study the problem of pliable private information retrieval with side information (PPIR-SI) for the single server case. In PPIR, the messages are partitioned into nonoverlapping classes and stored in a number of noncolluding databases. The user wishes to retrieve any one message from a desired class  while revealing no information about the desired class identity to the databases. In PPIR-SI, the user has prior access to some side information in the form of messages from different classes and wishes to retrieve any one \emph{new} message from a desired class, i.e., the message is not included in the side information set, while revealing no information about the desired class to the databases. We characterize the  capacity of (linear) single-server PPIR-SI for the case where the user's side information is \emph{unidentified}, i.e., the user is oblivious of the identities of its side information messages and the database structure. We term this case PPIR-USI. Surprisingly, we show that having side information, in PPIR-USI, is disadvantageous, in terms of the download rate, compared to PPIR.
\end{abstract}

\begin{IEEEkeywords}
	Capacity, information-theoretic privacy, pliable index coding, private information retrieval, side information.    
\end{IEEEkeywords}

\section{Introduction}
\label{sec:introduction}

In content delivery networks (CDNs), compromised network operators have the ability to monitor network traffic and deduce user preferences and the popularity of content. This information might be used maliciously to violate users' privacy and harm the content providers. As a result,  the protection of the delivered content \emph{type}, and thus the privacy of users' preferences and the popularity of the content, is an essential requirement of content delivery solutions \cite{CuiRizwanAsgharRussello20_1}.

Recently, inspired by privacy concerns in content-based applications, the problem of pliable private information retrieval (PPIR) was introduced \cite{ObeadKliewer22_1}. PPIR is a new variant of the well-established private information retrieval (PIR) problem that originated in theoretical computer science by Chor \emph{et.\ al} \cite{ChorGoldreichKushilevitzSudan98_1,Gasarch04_1}. 
In PIR, the goal is to efficiently retrieve a specific message from remote storage while keeping the identity of the desired message hidden. 
In contrast to the classical PIR problem, in PPIR, the messages are partitioned into several classes, and the user is flexible with her demand, i.e., the user is interested in privately retrieving \emph{any} one message from the messages belonging to a given class. 
Moreover, in PPIR, the definition of privacy differs from the classical PIR problem in terms of protecting the class identity instead of the desired message identity. This results in a trade-off between communication efficiency and the provided privacy level. 
Another related work,  known as digital blind box \cite{WangUlukus22_1}, considers a flexible variant of PIR where the user is also interested in retrieving any message, however, while keeping the identity of the message private from the remote storage.

We consider the problem of PPIR with side information (PPIR-SI). 
To motivate the problem, consider a content distribution network consisting of a server (content provider) with a large library of messages partitioned into several contents \emph{types}.
The network users share a broadcast medium and drop in and out of the network randomly. 
In many scenarios, users obtain information about/from the library a priori. 
For example, messages that have been broadcasted in a previous content update or pre-fetched during a caching stage. Each user stores a number of messages locally based on her preference.
This information is known as side information, and the problem of PIR with side information (PIR-SI) has received considerable attention (e.g., see \cite{WeiBanawanUlukus19_2, WeiUlukus20_1,  KazemiKarimiHeidarzadehSprintson19_1,LiGastpar20_2,KadheGarciaHeidarzadehElRouayhebSprintson20_1,ChenWangJafar20_2,HeidarzadehKazemiSprintson21_1,HeidarzadehSprintson22_1,Gomez-LeosHeidarzadeh22_1,LuJafar23_1,SiavoshaniShariatpanahiMaddah-Ali21_1}) due to its close ties with index coding (IC) \cite{BirkKol98}. IC is an active research area in 
network information theory that models problems in satellite communication, content broadcasting, and distributed storage and caching \cite{BirkKol06,Bar-YossefBirkJayramKol11_1,KarmooseSongCardoneFragouli17_1,ArbabjolfaeiKim18_1}. 
However, in the content distribution context, the goal is for the server to efficiently update the content of %
active users without prior knowledge of their local storage content and for the users to obtain new content without revealing their preferences. This problem is modeled by single-server PPIR with (unidentified) side information and is conceptually related to new variants of IC  (e.g., see \cite{BrahamaFragouli15_1,kaoMaddah-AliAvestimehr16_1}).

To the best of our knowledge, the problem of PPIR-SI has not been examined in the open literature. 
In this work, we formally define the PPIR-SI problem and address the following questions: Is side information helpful in reducing the communication cost compared to PPIR?
Can we achieve a better rate than for the parallel single-server PIR-SI problem \cite{KadheGarciaHeidarzadehElRouayhebSprintson20_1}, i.e., does pliability help? 
Thus, our goal is to demonstrate whether side information can be leveraged to construct efficient PPIR-SI schemes. Towards that, we focus on the applications where users have minimum knowledge about the database and their side information, i.e., unidentified side information (USI), and characterize the capacity of (linear) PPIR-USI (see \cref{thm:CapacityPPIR_USI}).
Specifically, we relate the problem of single-server PPIR-USI to a \emph{new} instance of the oblivious pliable index coding (OB-PICOD) problem \cite{BrahamaFragouli15_1}. We term this instance:\emph{ data shuffling constrained OB-PICOD with restricted side information} and leverage its relationship to PPIR-USI to prove the converse bound and the achievable rate of (linear) PPIR-USI.

\subsection{Notation}
\label{sec:notation}

We denote by $\Naturals$ the set of all positive integers, and for $a\in\Naturals$,
$[a]\eqdef\{1,2,\ldots,a\}$ and $[a:b]\eqdef\{a,a+1,\ldots,b\}$ for $a,b\in \Naturals$, $a \leq b$. 
A random variable (RV) is denoted by a capital Roman letter, e.g., $X$, while its realization is denoted by the corresponding small Roman letter, e.g., $x$. Vectors are boldfaced, e.g., $\vect{X}$ denotes a random vector and $\vect{x}$ denotes a deterministic vector. Random matrices are represented by bold sans serif letters, e.g., $\vmat{X}$, where $\mat{X}$ represents its realization. In addition, sets are denoted by calligraphic upper case letters, e.g., $\set{X}$. For a given index set $\set{S}$, we also write $\vmat{X}^\set{S}$ to represent $\bigl\{\vmat{X}^{(v)}\colon v\in\set{S}\bigr\}$.
$\HP{X}$ represents the entropy of $X$ and $\MI{X}{Y}$ the mutual information between $X$ and $Y$. 
We use the customary code parameters $[n,k]$ to denote a code $\code{C}$ of blocklength $n$ and dimension $k$ over the finite field $\Field_q$ of size $q$, where $q$ is a power of a prime number.
The linear span of a set of vectors $\set{X}=\{\vect{x}_1,\ldots,\vect{x}_a\}$, $a\in \Naturals$, is denoted by $\spn{\vect{x}_1,\ldots,\vect{x}_a}$ or $\espn{\set{X}}$.

We now proceed with a general description of the system model and problem statement of PPIR-SI.

\section{Definitions and Problem Statement}
\label{sec:definitions-problem}

\subsection{System Model}
\label{sec:system-model}

\begin{figure}[t!]
	\centering
	\includegraphics[draft=false,scale=0.174]{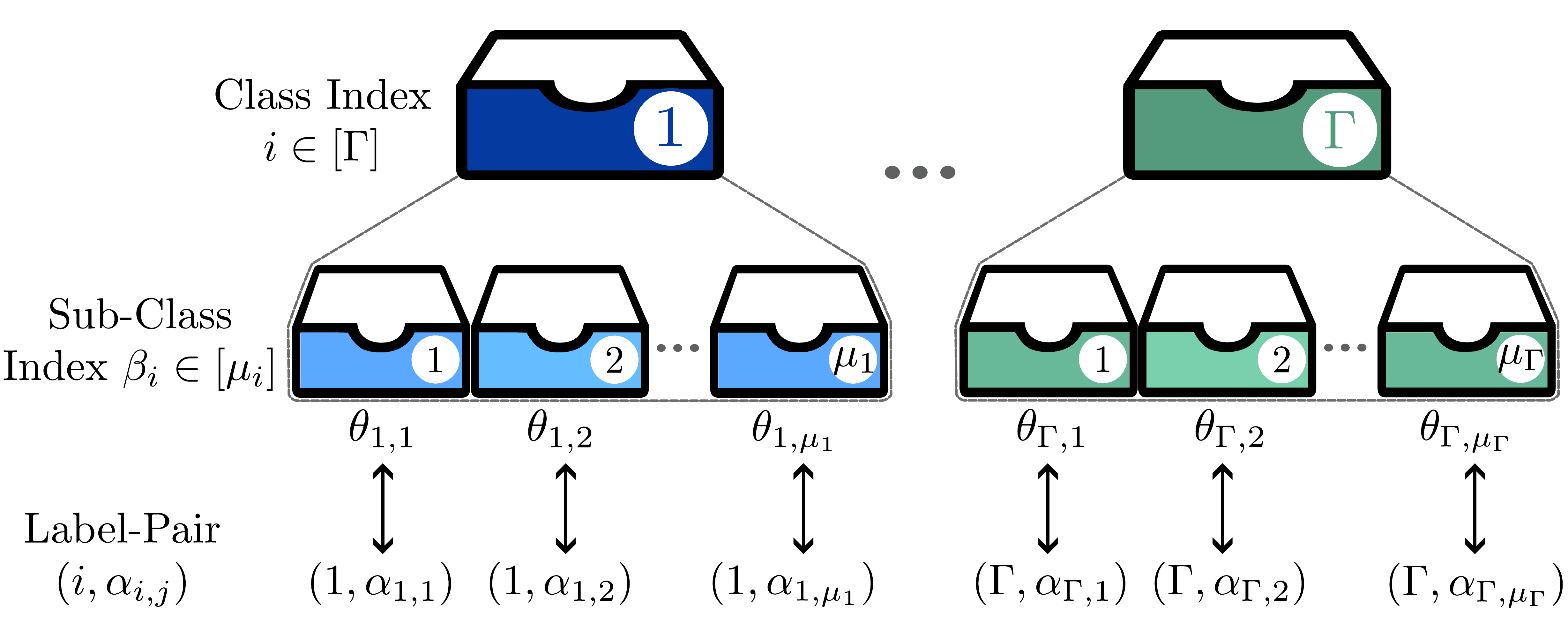}
	\caption{Index mapping of $f$ messages classified into $\Gamma$ classes using class and sub-class indices, i.e., $\theta_{i,\beta_{i}}\in {\mathcal M}_{i}\subset [f]$ for all $i\in[\Gamma]$. Each message index $\theta_{i,\beta_{i}}\in [f]$ is mapped to a label-pair $(i, \alpha_{i,j})$, where $\alpha_{i,j}\in \mathbb{N}$ is an identifier for each message selected independently of $f$ and $\mu_i$, for all $i\in[\Gamma]$ and $j\in [\mu_i]$.} 
	\label{fig:PPIR-IndexMapping}
\end{figure}

The single-server PPIR-SI problem is described as follows. We consider a database that stores in total $f$ independent
messages $\vect{W}^{(1)},\ldots,\vect{W}^{(f)}$, where each message $\vect{W}^{(m)}=\bigl(W_{1}^{(m)},\dots, W_{\const{L}}^{(m)}\bigr)$, $m\in [f]$, is a random length-$\const{L}$ vector with independent and identically distributed symbols that are chosen at random from the field $\Field_q$ for some
$\const{L} \in\Naturals$. The messages are partitioned or classified into $\Gamma$ classes where every message is classified into exactly one class, and no class is empty, i.e., $\Gamma \leq f$ and $\Gamma, f \in \mathbb{N}$. Moreover, we assume that there are at least two classes, i.e., $\Gamma\geq 2$. Without loss of generality, %
the symbols of each message are selected uniformly over the field  $\Field_q$. Thus,
$ \bigHP{\vect{W}^{(m)}} = \const{L}$, for all $m\in[f]$, and
$\bigHP{\vect{W}^{(1)},\dots,\vect{W}^{(f)}} = f\const{L}$ (in $q$-ary units).

Let $\set{M}_{i}$ be the set of \emph{message indices} belonging to the class indexed with $i\in[\Gamma]$, where $\mu_{i}\eqdef\card{\set{M}_{i}}$ is the size of this set.
Note that since every message is classified into exactly one class, then for all $i', i\in [\Gamma]$, where $i'\neq i$, we have $\set{M}_i \cap \set{M}_{i'} =\emptyset$, and $\sum_{i=1}^{\Gamma}\mu_i=f$. 
To represent the message index mapping that results from classifying the $f$ messages into $\Gamma$ classes, let, for $i\in[\Gamma]$, $\theta_{i,\beta_{i}}$ be the index of a message that belongs to class $i$ where $\beta_{i} \in [\mu_{i}]$ is a sub-class index and  $\theta_{i,\beta_{i}} \in \set{M}_{i}$. Here, the sub-class index $\beta_{i}$ represents the membership of a message \emph{within} the class $i$ as shown in Fig.~\ref{fig:PPIR-IndexMapping}. We assume that the message index mapping is kept private for the database and only a message label-pair is available for others, e.g., the label-pair is appended to the message itself. Thus, each message index $\theta_{i,\beta_{i}}\in [f]$ is mapped to a label-pair $(i, \alpha_{i,j})$, where $i\in [\Gamma]$ is the class index of the message and $\alpha_{i,j}\in \mathbb{N}$ is a random identifier selected independently of $f$ and $\mu_i$, however, uniquely within each class, i.e., $\alpha_{i, j}\neq \alpha_{i,j'}$ for $j, j' \in [\mu_i]$ and $j\neq j'.$

\subsection{Problem Statement} 
In single-server PPIR-SI, the user has access to $\kappa$ messages as side information. The messages are partitioned into $k_i\geq0$ messages from each class $i\in [\Gamma]$ and labeled with the label-pair set $ \hat{\set{S}} \triangleq \{ (1,\hat{\alpha}_{1,1}),\ldots, (1,\hat{\alpha}_{1,{k_1}}) , \ldots, (\Gamma,\hat{\alpha}_{\Gamma,{k_\Gamma}}) \}$,  where $\hat{\alpha}_{i,j} \in \Naturals$, for all $j\in [k_i]$, are identifiers unique within the same class.
Let $\collect{S}$ be the set of all unique subsets $\set{S}\subset [f]$ of size $|\set{S}|=\kappa$ that satisfy $\kappa=\sum_{i=1}^{\Gamma}k_i$ and there is exactly $k_i$ messages chosen from each class subset $\set{M}_i$ for all $i\in [\Gamma]$. The user, who is oblivious of the database structure, i.e., the indices of messages in each class, wishes to privately retrieve any one \emph{new} message $\vect{W}^{(m)}$ (if available) from a desired \emph{class}  $v\in [\Gamma]$, i.e., %
  $\vect{W}^{(m)}$ such that $m \in \set{M}_v$ and $m\notin \set{S}$ if $\set{M}_v\setminus\set{S}\neq \emptyset$,   
where $\set{S}\subset [f]$ is the index set corresponding to the user's side information. On the other hand, the database knows nothing about the user's side information set a priori.
\cref{tab:notation} summarizes important parameters. 
 \begin{table}[t!]  
 	\centering
 	\caption{Important Parameters} 
	\vspace{-2ex}
 	\label{tab:notation}
 	\Resize[0.9\columnwidth]{
 		\begin{IEEEeqnarraybox}[
 			\IEEEeqnarraystrutmode
 			\IEEEeqnarraystrutsizeadd{3pt}{2pt}]{v/c/V/s/v}
 			\IEEEeqnarrayrulerow\\
 			& \text{Notation} && Description\\ 	\hline\hline
 			& f                 &&  number of messages (integer)			      \\*\hline
            & \Gamma            &&  number of classes (integer) 		      	  \\*\hline  	
 			& \const{L}         &&  number of symbols in each message (integer)	  \\*\hline
            & \vect{W}^{(m)}    &&  $m$-th message (vector)                       \\*\hline
            & \set{M}_i         &&  set of message indices of class $i$ (set)     \\*\hline
            & \mu_i             &&  size of message indices set $\set{M}_i$ (integer)     \\*\hline
            & \kappa            && number of side information messages (integer)	\\*\hline
            & k_i               && number of side information messages from class $i$ (integer)
 			\\*\IEEEeqnarrayrulerow
 		\end{IEEEeqnarraybox}
 	}
 \end{table}

Let $S$, $\hat{S}$, $M$, and $V$ denote the RVs corresponding to the side information index set, the label-pair set at the user side, the message index, and the desired class index, respectively.
Following the classical PIR-SI literature, e.g., \cite{KazemiKarimiHeidarzadehSprintson19_1,LiGastpar20_2,KadheGarciaHeidarzadehElRouayhebSprintson20_1,ChenWangJafar20_2,HeidarzadehKazemiSprintson21_1,HeidarzadehSprintson22_1, Gomez-LeosHeidarzadeh22_1,LuJafar23_1}, we assume that the side information index set is uniformly distributed over all the sets in $\collect{S}$, i.e., $P_S(\set{S})=
\nicefrac{1}{\prod_{i=1}^{\Gamma} {\mu_i \choose k_i}}$. 
Note that, from the perspective of the database, there is a unique one-to-one mapping between the label-pair set %
and the side information index set, i.e., $\eHPcond{S}{\hat{S}}=0$. However, at this point, we do not assume that the indices of the side information messages $\set{S} \subset [f]$  are also available to the user. In other words, the mapping between $\set{S}$ and $\hat{\set{S}}$ is not given to the user.
The user privately selects a class index $v\in[\Gamma]$ and wishes to retrieve any \emph{new} message from the desired class while keeping the requested class index $v$ private from the server. We assume that the desired class index is uniformly distributed, i.e., $P_V(v)=\frac{1}{\Gamma}$.
It follows that\looseness-1
\begin{IEEEeqnarray*}{rCl}
P_M(m)=\begin{cases}
\frac{1}{\Gamma \mu_i} & \text{if } m \in \set{M}_i,\,\forall\,i\in [\Gamma], 
\\
0 & \text{otherwise}. 
\end{cases}
\end{IEEEeqnarray*}

Given side information, in order to retrieve a desired message $\vect{W}^{(m)}$ such that $m \in \set{M}_v$, $v\in[\Gamma]$, and $m\notin \set{S}$ if $\set{M}_v\setminus\set{S}\neq \emptyset$, from the database, the user sends a random query $Q^{(v,\set{S})}$ to the server. The query is generated by the user without any prior knowledge of the realizations of the stored messages. 
In other words, 
\begin{IEEEeqnarray*}{c}
\bigMI{\vect{W}^{(1)},\ldots,\vect{W}^{(f)}}{Q^{(V,S)}}=0. 
\end{IEEEeqnarray*}

In response to the received query, the server sends the answer $A^{(v,\set{S})}$ back to the user, where  $A^{(v,\set{S})}$ is a deterministic function of $Q^{(v,\set{S})}$ and the data stored in the database. Thus, 
$\bigHPcond{A^{(V,S)}}{Q^{(V,S)},\vect{W}^{[f]}}=0$ %
and $(S,V) \leftrightarrow (Q^{(V,S)},\vect{W}^{[f]}) \leftrightarrow A^{(V,S)}$ forms a Markov chain. 

To maintain user privacy, the query-answer function must be identically distributed for all possible class indices ${i\in[\Gamma]}$ from the perspective of the database. In other words, the scheme's queries and answer
strings must be independent of the desired class index. Moreover, the user must be able to reliably decode a \emph{new} message from the desired class $v$, i.e., $\vect{W}^{(m)}, $ where $m \in \mathcal{M}_v$ and $m\notin \set{S}$ if $\set{M}_v\setminus\set{S}\neq \emptyset$, using the received database answer and her side information $\set{S}$.

Consider a database storing $f$ messages classified into $\Gamma$ classes. The user has $\kappa=\sum_{i=1}^{\Gamma} k_i$ messages as side information, i.e., $ \vect{W}^{\hat{\set{S}}}$ where $ \hat{\set{S}} \triangleq \{ (1,\hat{\alpha}_{1,1}),\ldots, %
(\Gamma,\hat{\alpha}_{\Gamma,{k_\Gamma}}) \}$ and  $\hat{\alpha}_{i,j} \in \Naturals$ for all $i\in [\Gamma]$ and $j\in [k_i]$. For a desired class $v\in[\Gamma]$, the user wishes to retrieve any \emph{new} message $\vect{W}^{(m)}$,  where $m \in\mathcal{M}_v$ such that $m\notin\set{S}$ if $\set{M}_v\setminus\set{S}\neq \emptyset$, from query $Q^{(v,\set{S})}$ and answer $A^{(v, \set{S})}$. 
For a PPIR-SI protocol, %
there exists $m \in\mathcal{M}_v$ such that $m\notin\set{S}$ if $\set{M}_v\setminus\set{S}\neq \emptyset$, and the  recovery and privacy constraints, 
\begin{IEEEeqnarray}{rCl}    
  && \!\!\!\textnormal{[Recovery]}\; \bigHPcond{\vect{W}^{(m)}\!}{A^{(V,{S})}\!,Q^{(V,{S})}\!,\vect{W}^{\hat{{S}}}, V,S}=0,\label{eq:Recovery} 
  \\ 
  && \!\!\!\textnormal{[Privacy]}\quad\,
 \bigMI{V, S}{A^{(V,S)},Q^{(V,S)},\vect{W}^{[f]} }=0,\label{eq:Privacy}\IEEEeqnarraynumspace 
\end{IEEEeqnarray}%
must be satisfied.
Here, the privacy constraint represents a strong notion of privacy. Particularly, the server can not learn any information about both the desired class and the side information index set from the queries, the answers, and the stored messages, which is analogous to the so-called $(W,S)$-privacy definition in the PIR-SI problem \cite[Def.~2]{KadheGarciaHeidarzadehElRouayhebSprintson20_1}.
Finally, to measure the efficiency of a PPIR-SI protocol, we consider the required number of downloaded symbols for retrieving the
$\const{L}$ symbols of a \emph{new} message from a desired class. 
\begin{definition}[Rate and Capacity of Single-Server PPIR-SI]
	\label{def:PPIR-SIrate}
	The rate of a single-server PPIR-SI  protocol, denoted by $\const{R}_\mathrm{PPIR-SI}$, is defined as the ratio of the desired message size
	$\const{L}$ to the total required download cost $\const{D}$, i.e.,
	\begin{IEEEeqnarray*}{c}
		\const{R}_\mathrm{PPIR-SI}\eqdef\frac{\const{L}}{\const{D}}.
	\end{IEEEeqnarray*}
	The (linear) single-server PPIR-SI  capacity, denoted by $\const{C}_\mathrm{PPIR-SI}$, is the maximum achievable PPIR-SI rate over all possible (linear) PPIR-SI protocols. 
\end{definition}

\section{Main Results: Single-Server PPIR-USI}

The requirement of the user having full knowledge of the database structure might be infeasible in certain applications. In the PPIR-SI problem, we consider such applications and assume that the user is oblivious to the database structure. Moreover, since side information can  be obtained in several ways, 
e.g., the messages have been obtained previously through multi-message PPIR (M-PPIR) schemes \cite{ObeadKliewer22_1}, 
it is natural to also assume that the user has limited information about the messages in the side information set, e.g., the mapping from $\hat{\set{S}}$ to $\set{S}$ is not available at the user side.  
In this section, we assume that the user has very limited knowledge about her side information. 
Specifically, this case considers the following two restrictions:
1) the user does not know the identity of the side information messages $\set{S}\subset[f]$, but the user is aware of the class of each of the side information messages. 
2) The user does not know the size of each class $\mu_i$, for all $i\in [\Gamma]$, nor the size of the database $f$. 
We term this case PPIR with \emph{unidentified} side information (PPIR-USI) and add the natural assumption that $\mu_i\geq k_i+1$, for all $i\in [\Gamma]$. Namely, there is at least one message from every class that is not available at the user side, which in turn results in $\Gamma \leq f-\kappa $. Moreover, we restrict our discussion to linear PPIR-USI schemes.\footnote{ A linear scheme involves only linear operations over the messages, i.e.,  $A^{(v,\set{S})}$ is constructed from linear combinations of the $f$ messages \cite[Def.~1]{ZhangXu19_1}.} The main result is stated as follows. 

\begin{theorem}
\label{thm:CapacityPPIR_USI}
For the single-server PPIR-USI problem with $f$ messages, $\Gamma$ classes, and $\kappa$ unidentified side information messages portioned into $ \{k_i\}_{i\in[\Gamma]}$ out of $\{\mu_i \}_{i\in[\Gamma]} $ messages from each class, where $\mu_i\geq k_i+1$, the linear capacity is
    \begin{IEEEeqnarray}{rCl}    \label{eq:PPIR_USI} 
  \const{C}_\mathrm{PPIR-USI}=\inv{\left[\sum_{i=1}^{\Gamma} \min\{k_i+1, \mu_i-k_i \} \right]}. 
\end{IEEEeqnarray} 
\end{theorem}

The converse proof is provided in \cref{sec:USI-converse}, while in \cref{sec:USI-achievability} we devise an achievability scheme for PPIR-USI based on multiple maximum distance separable (MDS) codes.

\begin{remark}
\label{rem:1}
  It can be seen from \cref{thm:CapacityPPIR_USI} that the PPIR-USI capacity reduces to the capacity of single-server PPIR, i.e., $\const{C}_\mathrm{PPIR-USI}=\const{C}_\mathrm{PPIR}=\nicefrac{1}{\Gamma}$ \cite{ObeadKliewer22_1}, in two extreme points. Namely, when the user has no side information or the maximum amount of side information, i.e., $k_i\in \{0, \mu_i-1\}$ for all $i\in[\Gamma]$. Otherwise, %
  there is a penalty on the PPIR-USI rate to guarantee \emph{fresh} information is obtained from the desired class while leaking no information about its identity. Specifically, when $k_i+1\geq \mu_i-k_i$ for all $i\in [\Gamma]$, we can not do better than the capacity of single-server PIR-SI, i.e., $\const{C}_\mathrm{PPIR-USI}=\const{C}_\mathrm{PIR-SI}=\nicefrac{1}{(f-\kappa)}$ \cite{KadheGarciaHeidarzadehElRouayhebSprintson20_1}. Thus, the PPIR-USI capacity is bounded as $\nicefrac{1}{(f-\kappa)} \leq \const{C}_\mathrm{PPIR-USI} \leq  \nicefrac{1}{\Gamma}$.
\end{remark}

\begin{remark} 
\label{rem:2}
We say that the side information is \emph{identifiable} when more information regarding the side information and the database structure is available to the user, e.g., besides knowing the class of each message in the side information set, the user has the placement of each message within its ordered class. In this case, we conjecture that the rate %
 $\const{R}_\mathrm{PPIR-SI}=\nicefrac{1}{(\Gamma-\eta+1)}$ is achievable, where 
$\eta\leq \Gamma$ is the number of classes 
from which the user has identifiable side information.
\end{remark}

To support the claim of \cref{rem:2}, consider the simplest case where the user knows the size of each class. In this case, the user's side information set $\hat{\set{S}}=\{(i, \hat{\alpha}_{i,j}) \}_{i\in [\Gamma], j\in [k_i]}$ contains the placement of each message within its ordered class, i.e., $\hat{\alpha}_{i,j} \in [\mu_i]$. Here, the user queries the database with class and sub-class index pairs $\{ (1,\beta_{1}), \dots, (\Gamma,\beta_{\Gamma})\}$ and $\eta-1$ where $\eta-1$ indices of $\{\beta_i\}_{i\in[\Gamma]}$ are known to the user from $\hat{\set{S}}$ and the remaining $\Gamma-\eta+1$ indices of $\{\beta_i\}_{i\in[\Gamma]}$ are picked uniformly at random from $[\mu_i]$ (see the illustration of $\beta_i$ in \cref{fig:PPIR-IndexMapping}). 
Note that, if the side information set $\hat{\set{S}}$ includes messages from the desired class $v\in [\Gamma]$, then $\beta_v$ is picked uniformly at random from $[\mu_v]\setminus \{\hat{\alpha}_{v,j}\}_{j\in [k_v]}$. For the answer, the database encodes the messages corresponding to $\{ (1,\beta_{1}), \dots, (\Gamma,\beta_{\Gamma})\}$ with a systematic $[2\Gamma-\eta+1,\Gamma]$ MDS code and responds to the user with the $(\Gamma-\eta+1)\const{L}$ parity symbols. Accordingly, the achievable rate is equal to $\nicefrac{\const{L}}{(\Gamma-\eta+1)\const{L}}=\nicefrac{1}{(\Gamma-\eta+1)}$. We term this case, i.e., when the side information is \emph{fully} identifiable, as simply PPIR-FSI.

\begin{remark}
\label{rem:3}
For \cref{thm:CapacityPPIR_USI} to hold, $\mu_i \geq k_i+1$, for all $i\in [\Gamma]$, must be guaranteed. Otherwise, the PPIR-SI problem becomes challenging. Specifically, as the user is not aware of each class's size, she might query a class from which she possesses all available information, i.e.,  
there does not exist $\vect{W}^{(m)}$ such that $m\in \set{M}_i\setminus\set{S}$. Since any message from these classes is identifiable, we term this case PPIR with mixed side information (PPIR-MSI), and following \cref{rem:2}, we conjecture that the PPIR-MSI rate must be bounded as $\nicefrac{1}{(f-\kappa)} \leq \const{R}_\mathrm{PPIR-MSI}  \leq \nicefrac{1}{(\Gamma-\eta+1)}$.
Here, $\eta$, the number of classes from which the user has identifiable side information, becomes $\eta=\card{\{i \in [\Gamma]: \mu_i=k_i\}}$.
\end{remark}

Characterizing the capacities of PPIR-FSI and PPIR-MSI is left for future work. We now proceed with the converse proof of \cref{thm:CapacityPPIR_USI}.

\section{Converse Proof of \cref{thm:CapacityPPIR_USI}}
\label{sec:USI-converse}
The converse proof for the single-server PIR-SI capacity \cite{KadheGarciaHeidarzadehElRouayhebSprintson20_1} relies on showing that for any scheme that satisfies the required  recovery and privacy constraints, the answer from the server must be a solution to an instance of an IC problem that satisfies a number of requirements. 
Similarly, for the converse proof of \cref{thm:CapacityPPIR_USI}, we map the PPIR-USI problem to a \emph{new} instance of a data shuffling constrained OB-PICOD($t$) with restricted side information, where each client desires $t$ messages and all the clients possess the same number of side information messages. Then, we prove a lower bound on the broadcast rate of data shuffling constrained OB-PICOD($t$) with restricted side information for the case where $t\leq\Gamma$. Towards that, we first present the definition of the OB-PICOD($t$) problem \cite{BrahamaFragouli15_1}.  

\begin{definition}[OB-PICOD($t$)]
\label{def:OB-PICOD(t)}
Consider a server that has $f$ messages $\vect{W}^{[f]}\triangleq\{\vect{W}^{(1)} ,\dots, \vect{W}^{(f)} \}$, each with $\const{L}$ symbols from a finite field $\Field_q$ of size at least $q\geq 2f$, and $n$ clients $c_1,\dots,c_n$, $n\geq f$. Each client $c_j$, $j\in[n]$, has as side information a subset of messages $\vect{W}^{\set{S}_j}$, where $\set{S}_j \subset [f]$ is the set of indices of the messages that client $c_j$ has as side information and $|\set{S}_j|=\kappa$. Each client $c_j$ requests any $t$ messages $\{\vect{W}^{(m_{i,j})}\}_{i\in [t]}$, where $m_{i,j}\notin \set{S}_j$, for all $i\in [t]$, and $\kappa\leq f-t$. Given that the server has limited knowledge of the side information sets $\set{S}_j$, $j\in [n]$, i.e., only the size of the side information set $|\set{S}_j|=\kappa$ is known to the server, the (linear) OB-PICOD($t$) problem is to construct a linear code $\code{C}$ of length $l$ which consists of 
\begin{enumerate}
	\item a linear encoding function $\mathsf{Enc}: \Field_q^{f}\rightarrow\Field_q^{l}$ mapping the $\ell$-th symbols of the $f$ messages  $\vect{W}_\ell \eqdef\bigl(W^{(1)}_{\ell},W^{(2)}_{\ell},\ldots,W^{(f)}_{\ell}\bigr) \in \Field_q^{f}$, for all $ \ell\in [\const{L}]$, where $l$ is the length of the code. 
	\item Linear decoding functions $\mathsf{Dec}_j$ for all clients $c_j$, $j\in[n]$, such that $ \mathsf{Dec}_j (\{\mathsf{Enc}({\vect{W}_{\ell})}\}_{\ell\in[\const{L}]},\vect{W}^{\set{S}_j})=\{\vect{W}^{(m_{i,j})}\}_{i\in [t]}$ for some distinct $m_{i,j}\in [f]\setminus \set{S}_j$.
\end{enumerate}
\end{definition}

For the OB-PICOD($t$) problem, the optimal code length,  i.e., the minimum number of broadcast transmissions, is stated in  Theorem~\ref{thrm:OBPICOD}. 

\begin{theorem}[{\cite[Thm.~9]{BrahamaFragouli15_1}}] \label{thrm:OBPICOD}
	For any instance of OB-PICOD($t$) with $f$ messages, if each client has $\kappa$ messages as side information, then the minimum number of broadcasts needed to satisfy all of the clients is at most $\min\{\kappa+t,f-\kappa\}$ using linearly coded messages. 
\end{theorem}

The OB-PICOD$(t)$ problem does not directly map the problem formulation of PPIR-USI due to the specific structure of the side information sets and the privacy constraint. Thus, we introduce a new OB-PICOD$(t)$ instance. Specifically, we introduce a data shuffling constrained OB-PICOD$(t)$ with restricted side information sets. This new instance is inspired by the application of PICOD as a data shuffling approach in distributed computing \cite{SongFragouliZhao20_1} and is formally defined as follows.

\begin{definition}[Data Shuffling Constrained OB-PICOD($t$) With Restricted Side Information] \label{def:OB-PICOD(1)-RSI}
Consider an OB-PICOD($t$) problem consisting of a server with $f$ messages and $n$ clients $c_1, \dots, c_n$. 
Each client has $\kappa$ messages as side information and is interested in obtaining $t$ \emph{new} messages. However, the users' side information sets have a particular structure and the obtained messages must satisfy a data shuffling constraint. Specifically, for $ {i\in [\Gamma]}$, exactly $k_i\geq 0$ out of $\kappa$ side information messages are from the subset $\set{M}_i\subset [f]$ where $|\set{M}_i|=\mu_i$ and $\{\set{M}_i\}_{i\in [\Gamma]}$ are disjoint subsets satisfying $\bigcup_{i=1}^{\Gamma} \set{M}_i=[f]$. Moreover, the messages must be from any $t$ distinct subsets. Given that the server has limited knowledge of the side information sets $\set{S}_j$, $j\in [n]$, i.e., the restriction in each subset $\{k_i\}_{i\in[\Gamma]}$ and thus also the size of the side information sets $|\set{S}_j|=\kappa$ are known to the server, the (linear) data shuffling constrained OB-PICOD($t$) problem  with restricted side information is to construct a linear code $\code{C}$ of length $l$ with a linear encoding function as defined in \cref{def:OB-PICOD(t)} and linear decoding functions $\mathsf{Dec}_j$ for all clients $c_j$, $j\in[n]$, such that $ \mathsf{Dec}_{j} (\{\mathsf{Enc}({\vect{W}_{\ell})}\}_{\ell\in[\const{L}]},\vect{W}^{\set{S}_j})=\{\vect{W}^{(m_{i,j})}\}_{i\in \set{T}}$ for some distinct $m_{i,j}\in \set{M}_i\setminus \set{S}_j$, for arbitrary $\set{T}\subseteq{[\Gamma]}$ such that $\card{\set{T}}=t$. 
\end{definition}

We now map the data shuffling constrained OB-PICOD$(t)$ with restricted side information to our PPIR-USI problem.

\begin{lemma}
\label{lem:2}
For a single-server PPIR-USI scheme satisfying the privacy constraint of \eqref{eq:Privacy}, given a desired class index $v\in [\Gamma]$, a side information set $\set{S} \in \collect{S}$, and a query $Q^{(v, \set{S})}$, for every realization of messages $\vect{W}^{(1)}, \dots, \vect{W}^{(f)}$ divided into disjoint subsets $\{\set{M}_i\}_{i\in [\Gamma]}$, the answer $A^{(v, \set{S})}$ from the server must be a solution for the following instance of the data shuffling constrained  OB-PICOD$(\Gamma)$ with restricted side information. 
\begin{enumerate}
    \item The instance has the messages $\vect{W}^{(1)}, \dots, \vect{W}^{(f)}$.
    \item There are $n={\mu_1 \choose k_1} {\mu_2\choose k_2} \times \dots\times {\mu_\Gamma\choose k_\Gamma}$ clients with distinct side information sets consisting of $\kappa$ messages restricted to $k_i$ messages from subset $\set{M}_i$ for all $i\in[\Gamma]$.%
    \item  For every $\set{S}_j\in \collect{S}$, there exists a client $c_j$, $j\in [n]$, with side information $\set{S}_j$ that can decode a \emph{new} message from every subset $\set{M}_i$, $i \in [\Gamma]$. 
\end{enumerate}
\end{lemma}

The proof of \cref{lem:2} can be found in Appendix~\ref{app:lem2}. Note that, in \cref{lem:2}, the third condition represents the data shuffling constraint and guarantees that any client with  restricted side information set can decode at least one message from every subset. Thus, decoding at least $\Gamma$ \emph{new} messages in total. From a PPIR-USI perspective, this guarantees the privacy of the user's desired class jointly with the side information set. 
 
Finally, to conclude the converse proof of \cref{thm:CapacityPPIR_USI}, we solve the instance of OB-PICOD($\Gamma$) that satisfies the conditions specified in \cref{lem:2}. 
Although it is \emph{sufficient} to broadcast $\min\{\kappa+\Gamma,f-\kappa\}$  linearly coded messages to solve an instance of OB-PICOD($\Gamma$) with $f$ messages and side information sets of a fixed size $\kappa$ \cite[Lem.~10]{BrahamaFragouli15_1}, 
 the minimum number of broadcasts needed to satisfy all clients, i.e., the length of the linear OB-PICOD($\Gamma$) code, is at least
$$l=\sum_{i=1}^{\Gamma}\min \{k_i+1,\mu_i-k_i\}  \leq \min\{\kappa+\Gamma,f-\kappa\}$$  linearly coded messages. This result is stated in the following lemma. 


\begin{lemma} \label{lem:LowerBound-OBPICOD(1)}
Under a data shuffling constraint, any linear scheme for OB-PICOD($\Gamma$) with restricted side information subsets requires at least  $\sum_{i=1}^{\Gamma} \min \{k_i+1, \mu_i-k_i\}$ linearly coded messages.   
\end{lemma} 

The proving technique for Lemma~\ref{lem:LowerBound-OBPICOD(1)} follows from \cite[Lem.~11]{BrahamaFragouli15_1}, however, we tailor the proof to this specific instance of the data shuffling constrained OB-PICOD($t$) with restricted side information for $t\leq\Gamma$. The proof is found in Appendix~\ref{app:lem3}.

\noindent {\bf Converse Proof:}
For the single-server PPIR-USI problem with $f$ messages, $\Gamma$ classes, and $\set{S}\in \collect{S}$ side information set, it follows from Lemmas~\ref{lem:2} and \ref{lem:LowerBound-OBPICOD(1)} that the download cost is at least $\const{D}\geq l\const{L}$. Thus, the (linear) PPIR-USI capacity is at most $\const{C}_\mathrm{PPIR-USI} \leq \inv{\left[\sum_{i=1}^{\Gamma} \min \{k_i+1, \mu_i-k_i\} \right]}$.

\section{Achievability of \cref{thm:CapacityPPIR_USI}}
\label{sec:USI-achievability}
In this section, we propose a scheme for the single-server PPIR-USI problem, which achieves the rate of \eqref{eq:PPIR_USI} based on MDS codes. The scheme is described as follows.

\noindent {\bf PPIR-USI Scheme:} 
For a desired class index $v\in [\Gamma]$, the user queries the database with $ Q^{(v, \set{S})} = \{k_i\}_{i\in[\Gamma]}$, where $k_i$ is the size of side information from class $i\in [\Gamma]$.
Given $\{k_i\}_{i\in[\Gamma]}$, the database generates an answer as follows. 
For all $i\in [\Gamma]$,
\begin{itemize}
\item if $ k_i< \mu_i-k_i-1 $, randomly select $k_i+1$ messages from class $i$.
\item If $k_i \geq \mu_i-k_i-1$, encode across the messages of class $i$, i.e., encoding is over the $\ell$-th symbol of all messages in class $i$, $W^{(\theta_{i,1})}_\ell,\dots, W^{(\theta_{i,\mu_i})}_\ell$ for all $\ell \in[\const{L}]$ (see the illustration of $\theta_{i,\beta_i}$ in \cref{fig:PPIR-IndexMapping}), with a systematic $[2\mu_i-k_i,\mu_i]$ MDS code over the finite field $\Field_q$ with $q\geq 2\mu_i-k_i$. Then, select the $(\mu_i-k_i)\const{L}$ parity symbols. 
\end{itemize}
The selected messages are then transmitted as the answer $A^{(v,\set{S})}$ to the user.

\begin{lemma} \label{lem:achievable-PPIR-USI}
The PPIR-USI scheme satisfies the decodability and privacy constraints of \eqref{eq:Recovery} and \eqref{eq:Privacy}, respectively, and achieves a rate $\const{R}_\mathrm{PPIR-USI}= \nicefrac{1}{\sum_{i=1}^{\Gamma} \min\{k_i+1, \mu_i-k_i \}}$. 
\end{lemma}
\begin{IEEEproof}
\noindent {\bf Recovery:}
For class $i\in [\Gamma]$, if $k_i< \mu_i-k_i-1$, then the received $k_i+1$ messages are uncoded. Since the side information from this class is at most $k_i$, there exists \emph{at least one} received message that is not in the user side information set $\hat{\set{S}}$ and \emph{at most} $k_i+1$ messages. Next, if $k_i\geq \mu_i-k_i-1$, the received messages are encoded with a systematic $[2\mu_i-k_i,\mu_i]$ MDS code. Thus, given the $k_i$ side information messages and the received $\mu_i-k_i$ linear combinations of the $\mu_i$ messages, it follows that the user can recover the remaining $\mu_i-k_i$ messages. Finally, since $\mu_i\geq k_i+1$, the user is guaranteed to obtain \emph{at least one new} message from class $i\in [\Gamma]$ and \emph{at most} $\mu_i-k_i$ \emph{new} messages. Applying this argument for messages received from all classes $i\in[\Gamma]$, it can be easily seen that the user obtains \emph{at least one new} message from each class, including the desired class $v\in [\Gamma]$, i.e., $\Gamma$ messages in total, and \emph{at most} $\sum_{i=1}^{\Gamma} \min\{k_i+1, \mu_i-k_i \}\leq f-\kappa$ messages in total, thus satisfying the recovery condition of \eqref{eq:Recovery}.

\noindent {\bf Privacy:} Since the database only learns the sizes of the side information subsets $\{k_i\}_{i\in [\Gamma]}$ from the scheme's query, the query $Q^{(v, \set{S})}$ and answer $A^{(v,\set{S})}$ thus are independent of the particular realization of the desired class $v\in [\Gamma]$ and the side information index set $\set{S}\in \collect{S}$. Hence, the PPIR-USI scheme satisfies the privacy constraint \eqref{eq:Privacy}.

\noindent {\bf Achievable Rate:} To privately retrieve one \emph{new} message from the desired class, the user receives from every class $i\in [\Gamma]$ the minimum of $k_i+1$ uncoded messages {\m or} $\mu_i-k_i$ linearly encoded messages.
Following the fact that the $\mu_i$ messages in each class consist of $\const{L}$ symbols chosen independently and uniformly over the field $\Field_q$ and the $\mu_i-k_i$ encoded messages are a linear combination of these symbols, the symbols of the $\mu_i-k_i$ linear combinations are also independent and uniformly distributed over the field $\Field_q$.  Hence, we have
 $\const{D}=\HP{A^{(V,S)}}= \const{L}\sum_{i=1}^{\Gamma} \min\{k_i+1, \mu_i-k_i\}$ and, from \cref{def:PPIR-SIrate}, the achievable rate is $\const{R}_\mathrm{PPIR-USI}= \nicefrac{1}{\sum_{i=1}^{\Gamma} \min\{k_i+1, \mu_i-k_i \}}$. 
 \end{IEEEproof}

\begin{remark}
Following the achievable scheme for PPIR-USI, we highlight the following observations.
\begin{itemize}
\item  If $k_v=0$, i.e., we have no side information from the desired class, it may seem intuitive to discard the available side information from the undesired classes and achieve a rate of $\const{R}=\frac{1}{\Gamma}$ by simply invoking a PPIR scheme to obtain one message from the desired class. However, that is not a private solution for  PPIR-SI in general. 
Specifically, for a fixed side information set where $\set{B} = \{i : k_i > 0\}$, invoking a PPIR scheme \emph{depends} on the index of the desired class $v\notin \set{B}$, i.e., distinguishable queries, which violates the privacy constraint in \eqref{eq:Privacy}. In fact, the identity of the desired class is leaked completely if $\card{\set{B}}=1$.

\item The solution can be extended to the case where the user is interested in $\lambda$ messages from a desired class (or $\nu$ desired classes). For this case, it can be easily shown that the achievable rate becomes  $\const{R}_\mathrm{M-PPIR-USI}=\nicefrac{\lambda\nu}{\sum_{i=1}^{\Gamma} \min\{k_i+\lambda, \mu_i-k_i \}}$ under the condition that $\mu_i\geq k_i+\lambda$ for all $i\in [\Gamma]$. 

\end{itemize}
\end{remark}


\section{Conclusion}
In this work, we presented the problem of single-server PPIR-SI as a new variant of the PPIR problem.
For the scope of this work, we focused on the case where the user has very limited knowledge about the identity of the side information messages and the structure of the database. We termed this case as PPIR-USI and characterized the capacity. Unlike classical PIR-SI, having access to side information is not necessarily advantageous in terms of the information retrieval rate of PPIR-SI compared to PPIR. However, there is a trade-off between privacy and  communication cost when compared to PIR-SI.

\appendices 
 
\section{Proof of Lemma~\ref{lem:2}} 
\label{app:lem2}

Here, we prove that a solution for an instance of data shuffling constrained OB-PICOD($\Gamma$) with restricted side information that satisfies the conditions stated in \cref{lem:2} is also a solution for a PPIR-USI problem.

\begin{IEEEproof}
In a single-server PPIR-USI scheme, a user that desires a message from class $v\in [\Gamma]$ and has $\set{S}$ as side information set constructs the query $Q^{(v, \set{S})}$. For the scheme to satisfy the privacy constraint in \eqref{eq:Privacy}, the following statement must be true. For every side information set $\set{S}_j\in \collect{S}$, there exists a decoding function $ \mathsf{Dec}_{j}(A^{(v, \set{S})},\vect{W}^{\set{S}_j})=\{\vect{W}^{(m_{i,j})}\}_{i\in [\Gamma]}$ such that $m_{i,j}\in \set{M}_i\setminus\set{S}_j$. Otherwise, for a particular $(i,\set{S}_j)$, the database can deduce that a user with the side information set $\set{S}_j$ cannot decode a \emph{new} message from class $i$. As a result, $\HP{V,S}\neq \HPcond{V,S}{Q^{(V,S)},A^{(V,S)}}$ and the privacy condition of \eqref{eq:Privacy} is violated, i.e.,~$\MI{V, S}{ A^{(V,S)},Q^{(V,S)},\vect{W}^{[f]}}>0$.

Now, for a realization of the message set $\vect{W}^{[f]}$, let $A^{(v,\set{S})}$ be the answer of a single-server PPIR-USI scheme corresponding to the $Q^{(v,\set{S})}$ given $\vect{W}^{[f]}$. 
Then, for every $\set{S}_j\in\collect{S}$ a client with side information $\set{S}_j$ can use the linear decoding function $\mathsf{Dec}_{j}
(A^{(v, \set{S})},\vect{W}^{\set{S}_j})$ to decode one \emph{new} message from each $\set{M}_i$, i.e., $\Gamma$ \emph{new} messages in total. Thus, the answer $A^{(v,\set{S})}$ is also a solution to the data shuffling instance of OB-PICOD($\Gamma$) with restricted side information, as stated in the lemma. 
\end{IEEEproof}

\section{Proof of Lemma~\ref{lem:LowerBound-OBPICOD(1)}} \label{app:lem3}
In this appendix, we prove a lower bound on the length of a linear OB-PICOD($\Gamma$) code for restricted side information as stated in \cref{lem:LowerBound-OBPICOD(1)}.
\begin{IEEEproof} 
Consider a client with side information index set $\set{S} \in \collect{S}$, i.e., $|\set{S}|=\kappa=\sum_{i\in [\Gamma]}k_i$, where $\{k_i\}_{i\in [\Gamma]}$ is the number of side information messages from subset $\{\set{M}_i\}_{i\in [\Gamma]}$. Let $\vect{u}_i$ denote a unit vector in $\Field_q^f$, i.e., $\vect{u}_i$ has the identity element in the $i$-th position and zero otherwise. Thus, $\{ \vect{u}_i \}_{i\in [f]}$ is the  standard basis  of $\Field_q^f$. Moreover,  given the side information index set  $\set{S}$, the client can compute any vector in the $\spn{\vect{u}_{s_1},\dots, \vect{u}_{s_\kappa}}$ where $s_1, \dots , s_\kappa \in \set{S} \subseteq [f]$. Let $\set{U}_\set{S}$ denote the set of identity vectors corresponding to the side information messages, i.e., $ \set{U}_\set{S}= \{ \vect{u}_i \}_{i\in \set{S}}$. Since the server uses a length-$l$ linear code over  $\Field_q$ to encode the messages sent to the clients, we denote the encoding as 
\begin{IEEEeqnarray*}{rCl}
\trans{{\begin{pmatrix}
    \vect{W}^{(1)} \\
    \vect{W}^{(2)}  \\
    \vdots \\
    \vect{W}^{(f)} 
    \end{pmatrix}}_{f\times \const{L}}} {\begin{pmatrix}
    \vect{g}_1 & \vect{g}_2 & \dots & \vect{g}_l
    \end{pmatrix}}_{f\times l},
\end{IEEEeqnarray*}
where $\set{G}=\{ \vect{g}_1, \vect{g}_2, \dots, \vect{g}_{l}\}$ are the encoding (column) vectors, $\vect{W}^{(m)}$, $m\in [f]$, are length-$\const{L}$ message vectors, and $\trans{(\cdot)}$ denotes the transpose operator.

\begin{claim}[{\cite[Claim~1]{Bar-YossefBirkJayramKol11_1}}]
Client $c_j$ is able to decode a message indexed with $i\in [f]$ using its side information set corresponding to $\set{S}_j$ and the $l$ linearly encoded messages, if $\vect{u}_i$ belongs to $\spn{\set{G}\cup\set{U}_{\set{S}_j}}$. \label{clm:1}
\end{claim}

Before we start the proof, we present the used notation.
\begin{itemize}
\item For all $j\in[\Gamma]$, let $\rho_j=\min\{k_j+1,\mu_j-k_j\}$, the set $\set{P}\eqdef\{ \rho_{j_1}, \dots, \rho_{j_\Gamma}\}$ be the increasingly ordered set over $\{\rho_{j_i}\}_{j_i\in[\Gamma]}$, i.e., $\rho_{j_1} \leq \dots \leq \rho_{j_\Gamma}$, and $\set{J}_t=\{j_1,\dots,j_t \}$ be the set of indices of the first $t$ elements of ${\set{P}}$.

\item Let $\varrho_t=\sum_{j\in\set{J}_t} \rho_j$ and $\kappa_{t}=\sum_{j\in\set{J}_t}(k_j+1)$.
\item Let $\set{D}\subseteq [f]$ be the set of decoded message indices by all clients.
\end{itemize}

 For OB-PICOD($t$) with restricted side information where $t\leq\Gamma$, we first claim that the set of decoded messages by all clients $\set{D}$ satisfies $|\set{D}|\geq \kappa_t$.
If not, say $|\set{D}|< \kappa_t$, since every client has $\sum_{j\in\set{J}_t} k_j$ messages from the subset $\{\cup_{j\in\set{J}_t}\set{M}_{j}\}$, there exists a client with $\kappa$ messages as side information that can not decode $t$ \emph{new} messages. This contradicts  Definition~\ref{def:OB-PICOD(1)-RSI} of OB-PICOD($t$). 

Next, we prove that $\set{G}$ contains at least $\rho_t\leq \kappa_t$ linearly independent vectors.
For simplicity, assume that $|\set{D}|= \kappa_{t}$. 
Without loss of generality, assume that at least $k_j+1$ from the set of decoded messages $\set{D}$ are from the subset $\set{M}_j$ where $j\in \set{J}_t$, i.e., $\set{D}\subset \{\cup_{j\in \set{J}_t}\set{M}_j\}$. 
%
Let $\set{G}_{\set{D}}$ be the orthogonal projection of the encoding vectors $\set{G}$ onto the $\kappa_t$ coordinates corresponding to the set $\set{D}$, 
i.e., $\set{G}_{\set{D}}$ can be represented by
the column vectors of $ \mat{G}|_{\set{D}} $ 
where $\mat{G}$ is the $f\times l$ matrix whose columns are the encoding vectors $\set{G}= \{ \vect{g}_1, \dots, \vect{g}_l\}$ and ${\mat{G}}|_{\set{D}}$ denotes a $f\times l$ matrix whose rows $i\in \set{D}$ are the corresponding rows of $\mat{G}$ and each of the remaining rows $i\notin \set{D}$ is equal to the all-zero vector.
The following claim is a direct result of \cref{clm:1} and the definition of $\set{G}_{\set{D}}$. 

\begin{claim}\label{clm:2}  
Let $i\in[f]$ be the index of a message decoded by client $c_j$, i.e., $i\in \set{D}$ and, from \cref{clm:1},  $\vect{u}_i \in \spn{\set{G}\cup\set{U}_{\set{S}_j}}$. 
If $i\notin \set{S}_j$, then $\vect{u}_i$ must be orthogonal to $\set{U}_{\set{S}_j}$ and thus $\vect{u}_i \in\spn{\set{G}_{\set{D}}}$~\cite{BrahamaFragouli15_1}. 
\end{claim}

We now proceed with the main steps of the proof. 

\subsubsection{\textbf{Case 1.} $k_j+1 \leq \mu_j-k_j$ for all $j\in \set{J}_t$}
\label{sec:case-1}

In this case, we have
$\varrho_t=\kappa_t$. 
Then, we consider all possible overlapping patterns between $\set{D}$ and the side information index sets in $\collect{S}$.  
\begin{itemize}
\item First, since (i) $\ecard{\set{D}}=\kappa_t = \sum_{j\in \set{J}_t}(k_j+1)$; (ii) by assumption of this case we have $\mu_j\geq k_j+(k_j+1)$, for all $j\in \set{J}_t$; and (iii) from the problem statement we have $\mu_{j'}\geq k_{j'}$ for the remaining $j'\in [\Gamma]\setminus \set{J}_t$, one can select in total
  \begin{IEEEeqnarray*}{c}
    \kappa=\sum_{j\in\set{J}_t}k_j+\sum_{j'\in [\Gamma]\setminus \set{J}_t}k_{j'}
  \end{IEEEeqnarray*}
  side information indices 
  \begin{IEEEeqnarray*}{c}
    \set{S}_1 =\{s_{1,1}, s_{1,2}, \ldots, s_{1,\kappa}\}
  \end{IEEEeqnarray*}
  such that $\set{S}_1\cap\set{D}=\emptyset$.
Now, by combining the decodability constraint in \cref{def:OB-PICOD(1)-RSI} and \cref{clm:1}, for a client with the side information set $\set{S}_1$, at least one vector in $\{\vect{u}_{i}\}_{i\in \set{D}}$ must be in  $\spn{\set{G}\cup\set{U}_{\set{S}_1}}$. Let this vector be $\vect{u}_{\tau_1}$. It follows that $\tau_1\in \set{D}$. Moreover, since $\set{S}_1\cap \set{D}=\emptyset$, following \cref{clm:2}, $\vect{u}_{\tau_1}$ must be orthogonal to $\set{U}_{S_1}$ and thus $\vect{u}_{\tau_1}\in\spn{\set{G}_{\set{D}}}$.
  
\item Next, consider the side information set 
  \begin{IEEEeqnarray*}{rCl}
    \set{S}_2 =\{\tau_1, s_{2,1}, \dots, s_{2,\kappa-1}\},
  \end{IEEEeqnarray*}
  where $\set{S}_2\cap \set{D}=\{\tau_1\}$ and $\set{U}_{\set{S}_2}=\{\vect{u}_i \}_{i\in \set{S}_2}$. In this case, since the client, given the side information set $\set{S}_2$ and the linearly encoded messages, is able to decode at least one \emph{new} message, there must exist $\vect{u}_{\tau_2}\in\espn{\set{G}\cup\set{U}_{\set{S}_2}}$, i.e., $\tau_2\in \set{D}$ and following \cref{clm:2}, $\vect{u}_{\tau_2}\in\espn{\set{G}_{\set{D}}}$.

\item Now, given that any side information set is limited to $k_i$ messages from $\set{M}_i$, we repeat this decoding argument for 
  $\kappa_t-t+1$ side information set-\emph{types}. For the last set-\emph{type}, we consider the side information set 
  \begin{IEEEeqnarray*}{rCl}
    \set{S}_{\kappa_t-t+1} &= &\{\tau_1, \dots, \tau_{\kappa_t-t}, s_{\kappa_t-t+1,1}, \dots\\*
    &&\qquad \qquad\qquad  \dots, s_{\kappa_t-t+1,\kappa-\kappa_t+t}\},
  \end{IEEEeqnarray*}
  where $\set{S}_{\kappa_t-t+1}\cap \set{D}=\{\tau_1,\dots,\tau_{\kappa_t-t}\}$ and $\set{U}_{\set{S}_{\kappa_t-t+1}}=\{\vect{u}_i \}_{i\in \set{S}_{\kappa_t-t+1}}$. For this client to be able to decode a \emph{new} message, there must exist $\vect{u}_{\tau_{\kappa_t-t+1}} \in \spn{\set{G}\cup \set{U}_{\set{S}_{\kappa_t-t+1}}}$. Moreover, $\vect{u}_{\tau_{\kappa_t-t+1}}$  must be orthogonal to $\set{U}_{\set{S}_{\kappa_t-t+1}}$, thus $\vect{u}_{\tau_{\kappa_t-t+1}}\in\espn{\set{G}_{\set{D}}}$. Finally, since the user needs to decode $t$ \emph{new} messages, there must exist an additional $\vect{u}_{\tau_{\kappa_t-t+2}}, \dots, \vect{u}_{\tau_{\kappa_t}}\notin \set{U}_{\set{S}_{\kappa_t-t+1}} $ but in $ \espn{\set{G}\cup\set{U}_{\set{S}_{\kappa_t-t+1}}}$. Thus, ${\tau_{\kappa_t-t+2}}, \dots, {\tau_{\kappa_t}} \in \set{D}$ and $\vect{u}_{\tau_{\kappa_t-t+2}}, \dots, \vect{u}_{\tau_{\kappa_t}}\in\espn{\set{G}_{\set{D}}}$. 
\end{itemize}
Note that if for all $j\in \set{J}_t$, $k_j+1\leq \mu_j-k_j$, then any side information set $\set{S}\in\collect{S}$ falls in the definition of one of the set-\emph{types} $\{\set{S}_1, \dots, \set{S}_{\kappa_t-t+1}\}$, thus there always exists a message in $\set{D}$ that satisfies the decodability constraint for any client with side information set $\set{S}\in \collect{S}$. Furthermore, by definition, $\vect{u}_{\tau_1},\dots, \vect{u}_{\tau_{\kappa_t}}$ are all orthogonal and linearly independent, thus, $\vect{u}_{\tau_1},\dots, \vect{u}_{\tau_{\kappa_t}}\in\spn{\set{G}_{\set{D}}}$ implies that $\set{G}_{\set{D}}$ contains at least $\kappa_t$ linearly independent vectors. Finally, following the fact that an orthogonal projection preserves orthogonality, $\set{G}$ must also contain at least $\kappa_t=\rho_t$ orthogonal and linearly independent vectors. We now proceed by assuming the alternative case.

\subsubsection{\textbf{Case 2.} $\mu_j-k_j < k_j+1$ for some $j\in \set{J}_t$}
\label{sec:case-2}

In this case, $\varrho_t < \kappa_t$. Let $\tilde{\set{J}}\subseteq \set{J}_t$ where $\tilde{\set{J}}\eqdef \{j\in \set{J}_t: \mu_{j}-k_{j}< k_{j}+1\}$. Now, we consider all possible overlapping patterns between $\set{D}$ and the side information index sets in $\collect{S}$. 

First, consider a side information set $\tilde{\set{S}}$ that has the minimum overlap with the set of decoded messages $\set{D}$. Due to the fact that $|\set{D}|=\sum_{j\in\set{J}_t}(k_j+1)$ and $\mu_j-k_j<k_j+1$ for all $j\in \tilde{\set{J}}$, there exist exactly
\begin{IEEEeqnarray*}{c}
  \Delta\eqdef \sum_{j\in \tilde{\set{J}} } \Delta_j
\end{IEEEeqnarray*}
overlapping elements between $\tilde{\set{S}}$ and $\set{D}$, where $ \Delta_j \triangleq k_j+1 -(\mu_j-k_j)$ for $j\in \tilde{\set{J}}$, as illustrated in \cref{fig:OB-PICOD}. 
\begin{figure}[t!]
  \centering
  \includegraphics[draft=false,scale=0.34]{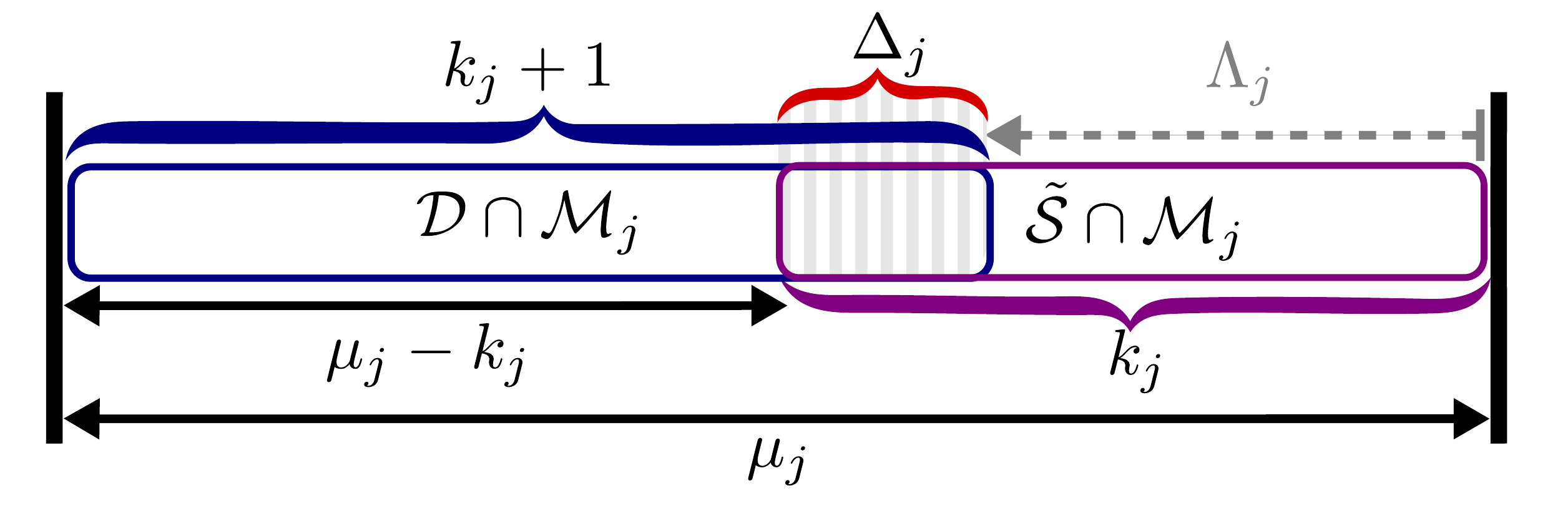}
  \caption{Minimum overlap between side information message set with the set of decoded messages $\set{D}$, where both sets are restricted to the subset $\set{M}_j$, i.e., $\set{D}\cap\tilde{\set{S}}\cap\set{M}_j$ for $j\in \tilde{\set{J}}$.}
  \label{fig:OB-PICOD}
\end{figure}
\begin{figure}[t!]
  \centering
  \includegraphics[draft=false,scale=0.34]{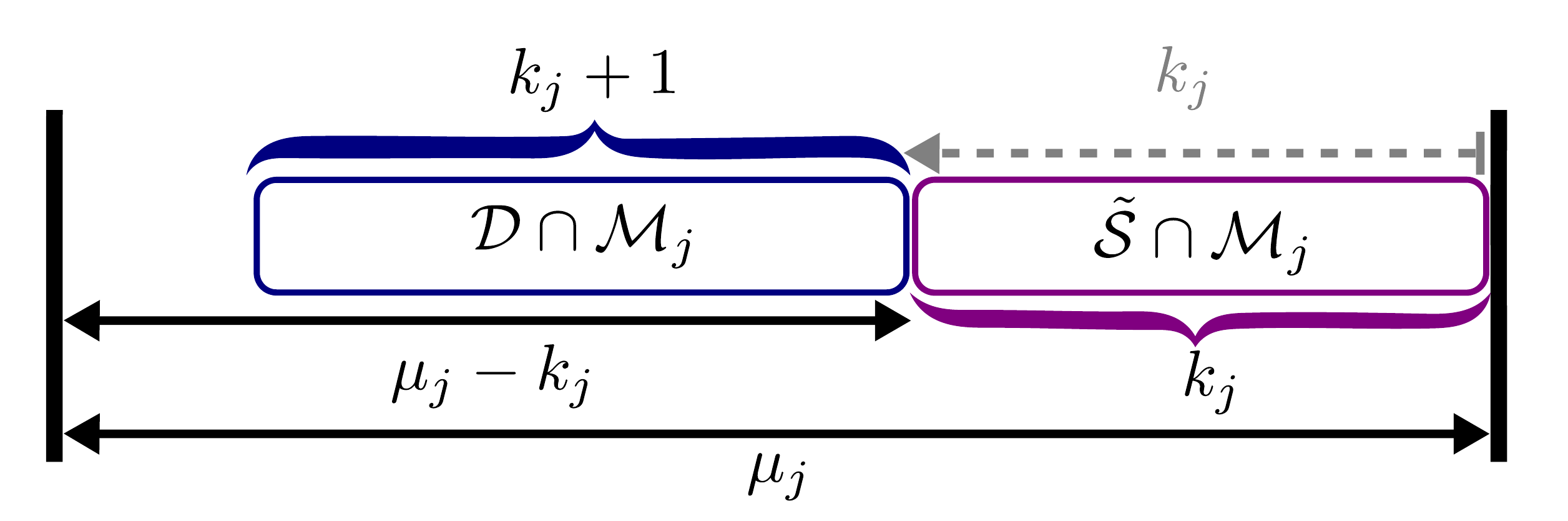}
  \caption{Minimum overlap between side information message set with the set of decoded messages $\set{D}$, where both sets are restricted to the subset $\set{M}_j$, i.e., $\set{D}\cap\tilde{\set{S}}\cap\set{M}_j$ for $j\in \set{J}_t\setminus\tilde{\set{J}}$.}
  \label{fig:OB-PICOD2}
\end{figure}

Thus, one can select
 \begin{IEEEeqnarray*}{c}
    \kappa=\underbrace{\sum_{j\in\tilde{\set{J}}} \Delta_j}_{\textnormal{overlap part}} +
    \sum_{j\in\tilde{\set{J}}} (k_j -\Delta_j) +
    \sum_{j\in \set{J}_t\setminus\tilde{\set{J}}} k_j +
    \sum_{j'\in [\Gamma]\setminus \set{J}_t}k_{j'}
  \end{IEEEeqnarray*}
side information indices
\begin{IEEEeqnarray*}{rCl}
\tilde{\set{S}}_{1} =\{\tau_1, \dots, \tau_{\Delta}, s_{1,1}, \dots, s_{1,\kappa-\Delta}\}
\end{IEEEeqnarray*}
such that $\tilde{\set{S}}_1\cap \set{D}= \{ \tau_1, \tau_2, \dots, \tau_{\Delta} \}$. Here, $\tilde{\set{S}}_{1}$ is the first side information set-\emph{type} with minimum overlap between side information message set and $\set{D}$.
%
Following the same decodability argument, for this client to be able to decode one \emph{new} message, there must exist $\vect{u}_{\hat{\tau}_{1}}\in\espn{\set{G}\cup\tilde{\set{S}}_{1}}$. Furthermore, $\vect{u}_{\hat{\tau}_{1}}$  must be orthogonal to $\set{U}_{\tilde{\set{S}}_1}$, thus $\vect{u}_{\hat{\tau}_{1}}\in \espn{\set{G}_{\set{D}}}$.

Now, repeat this step for the side information set-\emph{types} with elements from the subsets $\set{M}_j$ for $j\in \tilde{\set{J}}$ overlapping with the set of decoded messages $\set{D}$. Thus, we have $\Lambda$ additional side information set-\emph{types} $\tilde{\set{S}}_2, \dots, \tilde{\set{S}}_{\Lambda+1}$, where
\begin{IEEEeqnarray*}{c}
  \Lambda\eqdef\sum_{j\in \tilde{\set{J}}} \Lambda_j
\end{IEEEeqnarray*}
and $\ecard{\set{M}_j}-\ecard{\set{D}\cap\set{M}_j}=\Lambda_j\eqdef\mu_j-(k_j+1)$ for $j\in\tilde{\set{J}}$, as indicated in \cref{fig:OB-PICOD}. 

Next, we consider the side information set-\emph{type} with elements from the remaining subsets, i.e., $\set{M}_j$ for $j\in \set{J}_t\setminus\tilde{\set{J}}$, overlapping with the set $\set{D}$. As illustrated in \cref{fig:OB-PICOD2}, for the subsets $\set{M}_j$ where $j\in \set{J}_t\setminus \tilde{\set{J}}$, there are $k_j$ elements that can overlap with the set of decoded massages $\set{D}$. Thus, we have $\sum_{j\in \set{J}_t\setminus\tilde{\set{J}}} k_j$ additional side information set-\emph{types} $\tilde{\set{S}}_{\Lambda+2}, \dots,\tilde{\set{S}}_{\varrho_t-t+1}$, where 
\begin{IEEEeqnarray*}{rCl}
  \IEEEeqnarraymulticol{3}{l}{%
    (\Lambda+1)+  \sum_{j\in \set{J}_t\setminus\tilde{\set{J}}} \!\! k_j
  }\nonumber\\*\quad%
  & = &1+\sum_{j\in \tilde{\set{J}}} (\mu_j-k_j-1) +\sum_{j\in \set{J}_t\setminus\tilde{\set{J}}} k_j
  \\
  & = &\sum_{j\in \tilde{\set{J}}}(\mu_j-k_j) + \sum_{j\in\set{J}_t\setminus\tilde{\set{J}}}\!\! (k_j+1)-t+1
  \\
  & = &\sum_{j\in \set{J}_t}\min\{k_j+1,\mu_j-k_j\}-t+1
  \\
  & = &\varrho_t-t +1.
\end{IEEEeqnarray*}
Similar to \textbf{Case~1}, one can easily see that we will have $\vect{u}_{\hat{\tau}_{1}},\dots,  \vect{u}_{\hat{\tau}_{\varrho_t}}\in\espn{\set{G}_{\set{D}}}$ and by implication $ \set{G}_{\set{D}}$ and $\set{G}$ must contain at least $\varrho_t$ linearly independent vectors.

Finally, by combining the results of \textbf{Cases~1 and 2}, we conclude that  
$\set{G}$ must contain at least $\varrho_t$ linearly independent vectors. Therefore, the minimum number of broadcasts required to satisfy all OB-PICOD($t$) clients with restricted side information sets under a data shuffling constraint must be at least  $\varrho_t =\sum_{j\in \set{J}_t} \rho_j$. Subsequently, for $t=\Gamma$, it can be easily seen that $\varrho_\Gamma= \sum_{j\in [\Gamma]} \min\{k_j+1, \mu_j-k_j\}$.
\end{IEEEproof}
\IEEEtriggeratref{24}

\bibliographystyle{IEEEtran} 
\bibliography{./defshort1.bib, ./biblio1.bib,references.bib}

\end{document}